\documentstyle[psfig,12pt]{article}
\def\lsim{\:\raisebox{-0.5ex}{$\stackrel{\textstyle<}{\sim}$}\:}
\def\gsim{\:\raisebox{-0.5ex}{$\stackrel{\textstyle>}{\sim}$}\:}
\def\ip{in preparation}
\def\np#1#2#3{           {\it Nucl. Phys. }{\bf #1} (19#2) #3}
\def\be{\begin{equation}}       
\def\ee{\end{equation}}
\def\bear{\be\begin{array}}      
\def\eear{\end{array}\ee}
\def\bea{\begin{eqnarray}}
\def\eea{\end{eqnarray}}

\newcommand {\ignore}[1]{}

\newcommand{\bc}{\begin{center}}
\newcommand{\ec}{\end{center}}

\def\ifmath#1{\relax\ifmmode #1\else $#1$\fi}
\def\half{\ifmath{{\textstyle{1 \over 2}}}}
\def\quarter{\ifmath{{\textstyle{1 \over 4}}}}
\def\3quarter{{\textstyle{3 \over 4}}}
\def\third{\ifmath{{\textstyle{1 \over 3}}}}

\def\lf{\leaders\hbox to 1em{\hss.\hss}\hfill}
\def\21{$SU(2) \ot U(1)$}
\def\321{$SU(3) \ot SU(2) \ot U(1)$}
\def\ie{\hbox{\it i.e., }}

\def\eq#1{{eq. (\ref{#1})}}

\def\lsim{\raise0.3ex\hbox{$\;<$\kern-0.75em\raise-1.1ex\hbox{$\sim\;$}}}
\def\gsim{\raise0.3ex\hbox{$\;>$\kern-0.75em\raise-1.1ex\hbox{$\sim\;$}}}
\def\half{{1\over 2}}
\def\beq{\begin{equation}}
\def\eeq{\end{equation}}
\def\bef{\begin{figure}}
\def\eef{\end{figure}}
\def\bet{\begin{table}}
\def\eet{\end{table}}
\def\bea{\begin{eqnarray}}
\def\ba{\begin{array}}
\def\ea{\end{array}}
\def\bi{\begin{itemize}}
\def\ei{\end{itemize}}
\def\ben{\begin{enumerate}}
\def\een{\end{enumerate}}

\def\ot{\otimes}
\def\eea{\end{eqnarray}}
\def\np#1#2#3{           {\it Nucl. Phys. }{\bf #1} (19#2) #3}

\def\n.c.#1#2#3{         {\it Nuovo Cim. }{\bf #1} (19#2) #3}
\def\r.n.c.#1#2#3{       {\it Riv. del Nuovo Cim. }{\bf #1} (19#2) #3}

\def\ip{in preparation}
\def\ie{{\it i.e.}}

\def\half{{\textstyle{1 \over 2}}}
\def\third{{\textstyle{1 \over 3}}}
\def\quarter{{\textstyle{1 \over 4}}}
\def\sixth{{\textstyle{1 \over 6}}}
\def\eighth{{\textstyle{1 \over 8}}}
\def\bold#1{\setbox0=\hbox{$#1$}%
     \kern-.025em\copy0\kern-\wd0
     \kern.05em\copy0\kern-\wd0
     \kern-.025em\raise.0433em\box0 }
\begin{document}
%%%%%%%%%%%%%%%%%%%%%%%%%%% subequations.sty %%%%%%%%%%%%%%%%%%%%%%%%
\catcode`@=11
\newtoks\@stequation
\def\subequations{\refstepcounter{equation}%
\edef\@savedequation{\the\c@equation}%
  \@stequation=\expandafter{\theequation}%   %only want \theequation
  \edef\@savedtheequation{\the\@stequation}% % expanded once
  \edef\oldtheequation{\theequation}%
  \setcounter{equation}{0}%
  \def\theequation{\oldtheequation\alph{equation}}}
\def\endsubequations{\setcounter{equation}{\@savedequation}%
  \@stequation=\expandafter{\@savedtheequation}%
  \edef\theequation{\the\@stequation}\global\@ignoretrue

\noindent}
\catcode`@=12
%%%%%%%%%%%%%%%%%%%%%%%%%%%%%%%%%%%%%%%%%%%%%%%%%%%%%%%%%%%%%%%%%%%%%
\begin{titlepage}

\begin{flushright}
hep-ph/9706315\\
FTUV/97-32\\
IFIC/97-31\\
FISIST/8-97/CFIF\\
June 1997
\end{flushright}

\vspace*{5mm}

\begin{center}
{\Large \bf Minimal Supergravity with R--Parity Breaking}\\[15mm]

{\large{ Marco A. D\'\i az${}^1$, Jorge C. Rom\~ao${}^2$ and 
Jos\'e W. F. Valle${}^1$}\\
\hspace{3cm}\\
{\small ${}^1$Departamento de F\'\i sica Te\'orica, Universidad de Valencia}\\ 
{\small Burjassot, Valencia 46100, Spain}
\hspace{3cm}\\
\hspace{3cm}\\
{\small ${}^2$Departamento de F\'\i sica, Instituto Superior T\'ecnico}\\ 
{\small A. Rovisco Pais, 1096 Lisboa Codex, Portugal}} 
\end{center}
\vspace{5mm}
\begin{abstract}

We show that the minimal R-parity breaking model characterized by an
effective bilinear violation of R-parity in the superpotential is
consistent with minimal N=1 supergravity unification with radiative
breaking of the electroweak symmetry and universal scalar and gaugino
masses. This one-parameter extension of the MSSM-SUGRA model provides 
therefore the simplest reference model for the breaking of R-parity and
constitutes a consistent truncation of the complete dynamical models
with spontaneous R-parity breaking proposed previously. We comment on
the lowest-lying CP-even Higgs boson mass and discuss its minimal N=1
supergravity limit, determine the ranges of $\tan\beta$ and bottom
quark Yukawa couplings allowed in the model, as well as the relation
between the tau neutrino mass and the bilinear R-parity violating
parameter.

\end{abstract}

\end{titlepage}

%%%%%%%%%%%%%%%%%%%%%%%%%%%%%%%%%%%%%%%%%%%%%%%%%%%%%%%%%%%%%%%%%%%
\setcounter{page}{1}

\section{Introduction}

Supersymmetry apart from being attractive from the point of view of
providing a solution to the hierarchy problem and the unification of
the gauge couplings, provides an elegant mechanism for the breaking of
the electroweak symmetry via radiative corrections \cite{mssmrad}.  So
far most attention to the study of supersymmetric phenomenology has
been made in the framework of the Minimal Supersymmetric Standard
Model (MSSM) \cite{mssm} with conserved R-parity \cite{RP}.  R-parity
is a discrete symmetry assigned as $R_p=(-1)^{(3B+L+2S)}$, where L is
the lepton number, B is the baryon number and S is the spin of the
state. If R-parity is conserved all supersymmetric particles must
always be pair-produced, while the lightest of them must be stable.
Whether or not supersymmetry is realized with a conserved R-parity is
an open dynamical question, sensitive to physics at a more fundamental
scale.

The study of alternative supersymmetric scenarios where the effective
low energy theory violates R-parity \cite{HallSuzuki} has received 
recently a lot of attention \cite{mv90,beyond,RPothers}. As is well-known, 
the simplest supersymmetric extension of the Standard Model violates 
R-parity through a set of cubic superpotential terms involving a very 
large number of arbitrary Yukawa couplings. Although highly constrained 
by proton stability, one cannot exclude that a large number of such 
scenarios could be viable. Nevertheless their systematic study at a 
phenomenological level is hardly possible, due to the large number of 
parameters (almost fifty) characterizing these models, in addition to 
those of the MSSM.

As other fundamental symmetries, it could well be that R-parity is a
symmetry at the Lagrangean level but is broken by the ground state.
Such scenarios provide a very {\sl systematic} way to include R parity
violating effects, automatically consistent with low energy {\sl
baryon number conservation}. They have many added virtues, such as the
possibility of having a dynamical origin for the breaking of R-parity,
through radiative corrections, similar to the electroweak symmetry
\cite{rprad}.

In this paper we focus on the simplest truncated version of such a
model, in which the violation of R-parity is effectively parametrized
by a bilinear superpotential term $\epsilon_i\widehat L_i^a\widehat
H_2^b$. In this effective truncated model the superfield content is
exactly the standard one of the MSSM. In this case there is no
physical Goldstone boson, the Majoron, associated to the spontaneous
breaking of R-parity. Formulated at the weak scale, the model contains
only two new parameters in addition to those of the MSSM.
Alternatively, the unified version of the model, contains exactly a
single additional parameter when compared to the unified version of
the MSSM, which we will from now on call MSSM-SUGRA. Therefore our
model is the simplest way to break R-parity and can thus be regarded
as a reference model for R-parity breaking.  In contrast to models
with trilinear R-parity breaking couplings, it leads to a very
restrictive and systematic pattern of R-parity violating interactions.

Here we show that this simplest truncated version of the R-parity
breaking model of ref. \cite{rprad}, characterized by a bilinear
violation of R-parity in the superpotential, is consistent with
minimal N=1 supergravity models with radiative electroweak symmetry
breaking and universal scalar and gaugino masses at the unification
scale.  In particular, we perform a thorough study of the minimization
of the scalar boson potential using the tadpole method needed for an
accurate determination of the Higgs boson mass spectrum.  We comment
on the lowest-lying CP-even Higgs boson mass and discuss its minimal
N=1 supergravity limit, determining also the ranges of $\tan\beta$ and
bottom quark Yukawa couplings allowed at unification, as well as the
relation between the tau neutrino mass and the effective bilinear
R-parity violating parameter. Our results encourage further
theoretical work on this and on more complete versions of the model,
like that of ref. \cite{rprad}, as well as phenomenological studies of
the related signals.

\section{The Model}

The supersymmetric Lagrangian is specified by the superpotential $W$
given by \footnote{ We are using here the notation of
refs.~\cite{mssm} and \cite{GunHaber}.  }
\begin{equation}
W=\varepsilon_{ab}\left[
 h_U^{ij}\widehat Q_i^a\widehat U_j\widehat H_2^b
+h_D^{ij}\widehat Q_i^b\widehat D_j\widehat H_1^a
+h_E^{ij}\widehat L_i^b\widehat R_j\widehat H_1^a
-\mu\widehat H_1^a\widehat H_2^b
+\epsilon_i\widehat L_i^a\widehat H_2^b\right]
\label{eq:Wsuppot}
\end{equation}
where $i,j=1,2,3$ are generation indices, $a,b=1,2$ are $SU(2)$
indices, and $\varepsilon$ is a completely antisymmetric $2\times2$
matrix, with $\varepsilon_{12}=1$. The symbol ``hat'' over each letter
indicates a superfield, with $\widehat Q_i$, $\widehat L_i$, $\widehat
H_1$, and $\widehat H_2$ being $SU(2)$ doublets with hyper-charges
$\third$, $-1$, $-1$, and $1$ respectively, and $\widehat U$,
$\widehat D$, and $\widehat R$ being $SU(2)$ singlets with
hyper-charges $-{\textstyle{4\over 3}}$, ${\textstyle{2\over 3}}$, and
$2$ respectively. The couplings $h_U$, $h_D$ and $h_E$ are $3\times 3$
Yukawa matrices, and $\mu$ and $\epsilon_i$ are parameters with units
of mass. The first four terms in the superpotential are common to the
MSSM, and the last one is the only $R$--parity violating term. From
now on, we work only with the third generation of quarks and leptons.

Experimental evidence indicate that supersymmetry must be broken.
The actual supergravity mechanism is
unknown, but can be parametrized with a set of soft supersymmetry
breaking terms which do not introduce quadratic divergences to the
unrenormalized theory \cite{softterms}
\begin{eqnarray}
V_{soft}&=&
M_Q^{2}\widetilde Q^{a*}_3\widetilde Q^a_3+M_U^{2}
\widetilde U^*_3\widetilde U_3+M_D^{2}\widetilde D^*_3
\widetilde D_3+M_L^{2}\widetilde L^{a*}_3\widetilde L^a_3+
M_R^{2}\widetilde R^*_3\widetilde R_3\nonumber\\
&&\!\!\!\!+m_{H_1}^2 H^{a*}_1 H^a_1+m_{H_2}^2 H^{a*}_2 H^a_2-
\left[\half M_3\lambda_3\lambda_3+\half M_2\lambda_2\lambda_2
+\half M_1\lambda_1\lambda_1+h.c.\right]\label{eq:Vsoft} \\
&&\!\!\!\!\!\!\!\!\!\!\!\!\!\!\!\!\!\!\!\!+\varepsilon_{ab}\left[
A_th_t\widetilde Q^a_3\widetilde U_3 H_2^b
+A_bh_b\widetilde Q^b_3\widetilde D_3 H_1^a
+A_{\tau}h_{\tau}\widetilde L^b_3\widetilde R_3 H_1^a
-B\mu H_1^a H_2^b+B_2\epsilon_3\widetilde L^a_3 H_2^b\right]
\,.\nonumber
\end{eqnarray}
where we are already using a one--generation notation.

Note that in the effective low-energy supergravity model the bilinear
R-parity violating term {\sl cannot} be eliminated by superfield
redefinition even though it appears to be so at high scales, before
electroweak and supersymmetry breaking take place \cite{HallSuzuki}.
The reason is that the bottom Yukawa coupling, usually neglected in
the renormalization group evolution, plays a crucial role in splitting
the soft-breaking parameters $B$ and $B_2$ as well as the scalar
masses $m_{H_1}^2$ and $M_L^{2}$, assumed to be equal at the
unification scale. This can be seen explicitly from \eq{B} and \eq{B2}
as well as \eq{ML} and \eq{MHD} in Appendix A. This ensures that
R-parity violating effects can not be rotated away by going to a new
basis 
\footnote{Obviously physics does not depend on the choice of basis
\cite{DJV}. In this paper we choose to work with the unrotated fields.}
\cite{DJV,BabuJoshipura}, even if the starting RGE boundary
conditions for the soft-breaking terms are universal. 

It goes without saying that, in a supergravity model where
soft-breaking terms are not universal at the GUT scale, such as string
models, the bilinear violation of R-parity is also not removable.
However, in this case its effects are not calculable, in contrast to
our case. The same is true for the case of the most general low-energy
supersymmetric model \cite{eps0+}.

The electroweak symmetry is broken when the two Higgs doublets $H_1$
and $H_2$, and the tau--sneutrino acquire vacuum expectation values (VEVS):
\begin{eqnarray}
&&H_1={{{1\over{\sqrt{2}}}[\chi^0_1+v_1+i\varphi^0_1]}\choose{
H^-_1}}\,,\qquad
H_2={{H^+_2}\choose{{1\over{\sqrt{2}}}[\chi^0_2+v_2+
i\varphi^0_2]}}\,,
\nonumber \\
&&\qquad\qquad\qquad\widetilde L_3={{{1\over{\sqrt{2}}}
[\tilde\nu^R_{\tau}+v_3+i\tilde\nu^I_{\tau}]}\choose{\tilde\tau^-}}\,.
\label{eq:shiftdoub}
\end{eqnarray}
Note that the gauge bosons $W$ and $Z$ acquire masses given by
$m_W^2=\quarter g^2v^2$ and $m_Z^2=\quarter(g^2+g'^2)v^2$, where
$v^2\equiv v_1^2+v_2^2+v_3^2=(246 \; {\rm GeV})^2$.  We introduce the
following notation in spherical coordinates:
\begin{eqnarray}
v_1&=&v\sin\theta\cos\beta\cr
v_2&=&v\sin\theta\sin\beta\cr
v_3&=&v\cos\theta
\label{eq:vevs}
\end{eqnarray}
which preserves the MSSM definition $\tan\beta=v_2/v_1$. The angle $\theta$
equal to $\pi/2$ in the MSSM limit.

The full scalar potential may be written as
\begin{equation}
V_{total}  = \sum_i \left| { \partial W \over \partial z_i} \right|^2
	+ V_D + V_{soft} + V_{RC}
\label{V}
\end{equation}
where $z_i$ denotes any one of the scalar fields in the
theory, $V_D$ are the usual $D$-terms, $V_{soft}$ the SUSY soft
breaking terms given in eq.~(\ref{eq:Vsoft}), and $V_{RC}$ are the 
one-loop radiative corrections. It is popular to treat radiative 
corrections with the effective potential. In this case, 
$V_{RC}$ corresponds to the one--loop contributions to the effective
potential. Here we prefer to use the diagrammatic method and find 
the minimization conditions by correcting to one--loop the tadpole
equations. At the level of finding the minima, the two methods are
equivalent \cite{diazhaberii}. Nevertheless, the diagrammatic (tadpole)
method has advantages with respect to the effective potential when
we calculate the one--loop corrected scalar masses \cite{Vanderbilt}.

The scalar potential contains linear terms
\begin{equation}
V_{linear}=t_1^0\chi^0_1+t_2^0\chi^0_2+t_3^0\tilde\nu^R_{\tau}\,,
\label{eq:Vlinear}
\end{equation}
where
\begin{eqnarray}
t_1^0&\hskip -5pt=&\hskip -5pt(m_{H_1}^2+\mu^2)v_1-B\mu v_2-\mu\epsilon_3v_3+
\eighth(g^2+g'^2)v_1(v_1^2-v_2^2+v_3^2)\,,
\nonumber \\
t_2^0&\hskip -5pt=&\hskip -5pt(m_{H_2}^2+\mu^2+\epsilon_3^2)v_2-B\mu v_1+
B_2\epsilon_3v_3-\eighth(g^2+g'^2)v_2(v_1^2-v_2^2+v_3^2)\,,
\label{eq:tadpoles} \\
t_3^0&\hskip -5pt=&\hskip -5pt(m_{L_3}^2+\epsilon_3^2)v_3-\mu\epsilon_3v_1+
B_2\epsilon_3v_2+\eighth(g^2+g'^2)v_3(v_1^2-v_2^2+v_3^2)\,.
\nonumber
\end{eqnarray}
These $t_i^0, i=1,2,3$ are the tree level tadpoles, and are equal to 
zero at the minimum of the potential. 

\section{Squark Sector and Radiative Corrections}

To find the correct electroweak symmetry breaking radiatively, we need
to relate parameters at the GUT scale with parameters at the weak
scale. This means we are promoting the parameters in the tree level
tadpoles in eq.~(\ref{eq:tadpoles}) to running parameters. Therefore,
in order to find the correct minima of the scalar potential it is
essential to include the one--loop contributions to the tadpoles,
otherwise our tadpoles would be extremely scale dependent, \ie,
unphysical.

The main one--loop contributions to the tadpoles come from loops
involving top and bottom quarks and squarks. Therefore, we need to
study the scalar quark sector, and in particular, the spectrum and
couplings to CP--even neutral scalars.

The term $\epsilon_3\widehat L_3^a\widehat H_2^b$ in the
superpotential induce F--terms in the scalar potential, leading to
squark mass terms of the form $\tilde t_L\tilde t_R^*$ proportional to
$\epsilon_3$. In addition, the non--zero value of the vacuum
expectation value of the tau--sneutrino generates, from the D--terms,
squark mass terms of the form $\tilde t_i\tilde t_i^*$ and $\tilde
b_i\tilde b_i^*$, $i=L,R$. The new squark mass matrices are:
\begin{equation}
{\bold M}^2_{\tilde t}=\left[\matrix{
M_Q^2+m_t^2+\eighth(g^2-\third g'^2)(v_1^2-v_2^2+v_3^2) & 
m_t(A_t-\mu v_1/v_2+\epsilon_3 v_3/v_2) \cr
m_t(A_t-\mu v_1/v_2+\epsilon_3 v_3/v_2) &
M_U^2+m_t^2+\sixth g'^2(v_1^2-v_2^2+v_3^2)}\right]
\label{eq:tsquarkmm}
\end{equation}
for the top squarks, and
\begin{equation}
{\bold M}^2_{\tilde b}=\left[\matrix{
M_Q^2+m_b^2-\eighth(g^2+\third g'^2)(v_1^2-v_2^2+v_3^2) & 
m_b(A_b-\mu v_2/v_1) \cr m_b(A_b-\mu v_2/v_1) &
M_D^2+m_b^2-{\textstyle{1\over{12}}}g'^2(v_1^2-v_2^2+v_3^2)}\right]
\label{eq:bsquarkmm}
\end{equation}
for the bottom squarks. The reader can recover the MSSM squark mass
matrices by taking $\epsilon_3=v_3=0$ in the above two equations. The
quark masses are related to the quark Yukawa couplings in the same way
as in the MSSM: $m_t=h_tv_2/\sqrt{2}$ and
$m_b=h_bv_1/\sqrt{2}$. Nevertheless, the numerical value of the quark
Yukawas is higher in comparison with the MSSM to compensate with
smaller vacuum expectation values
\begin{equation}
h_t={{gm_t}\over{\sqrt{2}m_Ws_{\beta}s_{\theta}}}\,,\qquad
h_b={{gm_b}\over{\sqrt{2}m_Wc_{\beta}s_{\theta}}}\,,
\label{eq:quarkyukawas}
\end{equation}
and this is represented by the term $\sin\theta\equiv s_{\theta}$ in the 
denominators in the above equations.

Squark mass matrices ${\bold M}^2_{\tilde t}$ and ${\bold M}^2_{\tilde b}$
are diagonalized by two rotation matrices such that:
\begin{equation}
{\bold R}_{\tilde t}{\bold M}^2_{\tilde t}{\bold R}_{\tilde t}^T=
\left[\matrix{m_{\tilde t_1} & 0 \cr 0 & m_{\tilde t_2}}\right]\,,\qquad
{\bold R}_{\tilde b}{\bold M}^2_{\tilde b}{\bold R}_{\tilde b}^T=
\left[\matrix{m_{\tilde b_1} & 0 \cr 0 & m_{\tilde b_2}}\right]\,,
\label{eq:rotsqm}
\end{equation}
where $m_{\tilde q_1}<m_{\tilde q_2}$ by convention. These rotation
matrices play an important role in the determination of the scalar
couplings to a pair of squarks.

We introduce the notation for the CP--even neutral scalars 
$S^0_i=\chi^0_1,\chi^0_2,\tilde\nu_{\tau}^R$ for $i=1,2,3$ respectively. 
In this way, the Feynman rules of the type $S^0_i q \overline{q}$ are
\begin{equation}
\chi^0_1 b \overline{b}\longrightarrow -i{1\over{\sqrt{2}}}h_b\,,\qquad
\chi^0_2 t \overline{t}\longrightarrow -i{1\over{\sqrt{2}}}h_t\,.
\label{eq:S0qqrules}
\end{equation}
as in the MSSM, but with the quark Yukawa couplings given by
eq.~(\ref{eq:quarkyukawas}). Feynman rules of the type $S^0_i q
\overline{q}$ not listed in eq.~(\ref{eq:S0qqrules}) are zero.

In a similar way, we find Feynman rules of the type $S^0_i \tilde q
\tilde q^*$, \ie, CP--even neutral scalars couplings to a pair of
squarks. We start with $\chi^0_1$ couplings to top squarks:
\begin{eqnarray}
&&\chi^0_1\tilde t\tilde t^*\longrightarrow i{\bold M}_{\chi^0_1\tilde 
t\tilde t}\,,\qquad {\bold M}_{\chi^0_1\tilde t\tilde t}={\bold R}_{\tilde t}
{\bold M'}_{\chi^0_1\tilde t\tilde t}{\bold R}_{\tilde t}^T\,,
\nonumber\\
&&{\bold M'}_{\chi^0_1\tilde t\tilde t}=\left[\matrix{
-\quarter(g^2-\third g'^2)v_1 & {1\over{\sqrt{2}}}h_t\mu \cr
{1\over{\sqrt{2}}}h_t\mu & -\third g'^2v_1}\right]
\label{eq:S_1ststrules}
\end{eqnarray}
and to bottom squarks:
\begin{eqnarray}
&&\chi^0_1\tilde b\tilde b^*\longrightarrow i{\bold M}_{\chi^0_1\tilde 
b\tilde b}\,,\qquad {\bold M}_{\chi^0_1\tilde b\tilde b}={\bold R}_{\tilde b}
{\bold M'}_{\chi^0_1\tilde b\tilde b}{\bold R}_{\tilde b}^T\,,
\nonumber\\
&&{\bold M'}_{\chi^0_1\tilde b\tilde b}=\left[\matrix{
-h_b^2v_1+\quarter(g^2+\third g'^2)v_1 & -{1\over{\sqrt{2}}}h_bA_b \cr
-{1\over{\sqrt{2}}}h_bA_b & -h_b^2v_1+\sixth g'^2v_1}\right]
\label{eq:S_1sbsbrules}
\end{eqnarray}
These couplings have the same form in the MSSM but, as it was said before, the
Yukawa couplings are different and given by eq.~(\ref{eq:quarkyukawas}). In 
addition, vacuum expectation values $v_1$ and $v_2$ are different with respect
to the MSSM and given by $v_1=2m_Wc_{\beta}s_{\theta}/g$ and
$v_2=2m_Ws_{\beta}s_{\theta}/g$ and again, the deviation from the MSSM is
parametrized by the angle $\theta$.

Now we turn to the neutral CP-even Higgs $\chi^0_2$ that comes from
the second Higgs doublet. Its couplings to top squarks are:
\begin{eqnarray}
&&\chi^0_2\tilde t\tilde t^*\longrightarrow i{\bold M}_{\chi^0_2\tilde 
t\tilde t}\,,\qquad {\bold M}_{\chi^0_2\tilde t\tilde t}={\bold R}_{\tilde t}
{\bold M'}_{\chi^0_2\tilde t\tilde t}{\bold R}_{\tilde t}^T\,,
\nonumber\\
&&{\bold M'}_{\chi^0_2\tilde t\tilde t}=\left[\matrix{
-h_t^2v_2+\quarter(g^2-\third g'^2)v_2 & -{1\over{\sqrt{2}}}h_tA_t \cr
-{1\over{\sqrt{2}}}h_tA_t & -h_t^2v_2+\third g'^2v_2}\right]
\label{eq:S_2ststrules}
\end{eqnarray}
and to bottom squarks:
\begin{eqnarray}
&&\chi^0_2\tilde b\tilde b^*\longrightarrow i{\bold M}_{\chi^0_2\tilde 
b\tilde b}\,,\qquad {\bold M}_{\chi^0_2\tilde b\tilde b}={\bold R}_{\tilde b}
{\bold M'}_{\chi^0_2\tilde b\tilde b}{\bold R}_{\tilde b}^T\,,
\nonumber\\
&&{\bold M'}_{\chi^0_2\tilde b\tilde b}=\left[\matrix{
-\quarter(g^2+\third g'^2)v_2 & {1\over{\sqrt{2}}}h_b\mu \cr
{1\over{\sqrt{2}}}h_b\mu & -\sixth g'^2v_2}\right]
\label{eq:S_2sbsbrules}
\end{eqnarray}

Finally, we turn to the real part of the tau--sneutrino field, which mixes
with $\chi^0_1$ and $\chi^0_2$. Its couplings to top squarks are:
\begin{eqnarray}
&&\tilde\nu_{\tau}^R\tilde t\tilde t^*\longrightarrow i{\bold M}_{\tilde\nu_
{\tau}^R\tilde t\tilde t}\,,\qquad {\bold M}_{\tilde\nu_{\tau}^R\tilde t
\tilde t}={\bold R}_{\tilde t}{\bold M'}_{\tilde\nu_{\tau}^R\tilde t
\tilde t}{\bold R}_{\tilde t}^T\,,
\nonumber\\
&&{\bold M'}_{\tilde\nu_{\tau}^R\tilde t\tilde t}=\left[\matrix{
-\quarter(g^2-\third g'^2)v_3 & -{1\over{\sqrt{2}}}h_t\epsilon_3 \cr
-{1\over{\sqrt{2}}}h_t\epsilon_3 & -\third g'^2v_3}\right]
\label{eq:S_3ststrules}
\end{eqnarray}
and to bottom squarks:
\begin{eqnarray}
&&\tilde\nu_{\tau}^R\tilde b\tilde b^*\longrightarrow i{\bold M}_{\tilde\nu_
{\tau}^R\tilde b\tilde b}\,,\qquad {\bold M}_{\tilde\nu_{\tau}^R\tilde b
\tilde b}={\bold R}_{\tilde b}{\bold M'}_{\tilde\nu_{\tau}^R\tilde b
\tilde b}{\bold R}_{\tilde b}^T\,,
\nonumber\\
&&{\bold M'}_{\tilde\nu_{\tau}^R\tilde b\tilde b}=\left[\matrix{
\quarter(g^2+\third g'^2)v_3 & 0 \cr 0 & \sixth g'^2v_3}\right]
\label{eq:S_3sbsbrules}
\end{eqnarray}
These couplings $\tilde\nu_{\tau}^R \tilde q \tilde q^*$ vanish in the MSSM 
limit $v_3=\epsilon_3=0$, as it should.
 
We are now ready to include the effect of the one--loop tadpoles in
eq.~(\ref{eq:tadpoles}). The first step towards the calculation of radiative
corrections is the introduction of counter-terms. All parameters in the
Lagrangian are shifted from bare parameters to renormalized parameters
minus a counter-term:
\begin{eqnarray}
\lambda \longrightarrow & \lambda-\delta\lambda &\qquad 
\lambda=g,g',h_t,h_b,h_{\tau}, 
\nonumber\\
m^2 \longrightarrow & m^2-\delta m^2 &\qquad m^2=m_{H_1}^2,m_{H_2}^2,
m_{L_3}^2,m_{R_3}^2,\mu,\epsilon_3,
\nonumber\\
v_i \longrightarrow & v_i-\delta v_i &\qquad i=1,2,3,
\label{eq:bareshift}\\
A \longrightarrow & A-\delta A &\qquad A=A_t,A_b,A_{\tau},
\nonumber\\
B \longrightarrow & B-\delta B &\qquad B=B,B_2,
\end{eqnarray}
for couplings, masses, vacuum expectation values, trilinear soft
parameters, and bilinear soft parameters respectively.  If we make
this shift in the tadpole equations given in eq.~(\ref{eq:tadpoles}),
the tadpole themselves get a counter-term $\delta t_i$ for
$i=1,2,3$. Therefore, the one--loop tadpole equations are 
\begin{equation} 
t_i=t_i^0-\delta t_i+T_i(Q)\,,\qquad i=1,2,3,
\label{eq:RenTadpole} 
\end{equation} 
where $t_i$ are the one-loop
renormalized tadpoles and $T_i(Q)$ are the one--loop contributions to
the tadpoles, with $Q$ being the arbitrary mass scale introduced by
Dimensional Reduction.

The renormalization scheme we choose to work with is the 
$\overline{MS}$ scheme, where by definition the tadpole counter-terms are 
taken such that they cancel the divergent pieces of $T_i(Q)$ proportional
to $\Delta$:
\begin{equation}
\Delta={2\over{4-n}}+\ln 4\pi-\gamma_E,
\label{eq:Delta}
\end{equation}
where $\Delta$ is the regulator of dimensional regularization, $n$ is the
number of space--time dimensions, and $\gamma_E$ is the Euler's constant.
The $\overline{MS}$--counter-terms chosen in this way make the tadpoles
finite. We introduce the notation
\begin{equation}
\widetilde T_i^{\overline{MS}}(Q)=-\delta t_i^{\overline{MS}}+T_i(Q),
\label{eq:finite1loopTad}
\end{equation}
for the finite one--loop contribution to the tadpoles. These finite one--loop
tadpoles depend explicitly on the arbitrary scale $Q$. 

The one--loop tadpoles $t_i$ must be scale independent (at least in the
one--loop approximation), therefore, the renormalized parameters are
promoted to running parameters, \ie, they evolve with the scale $Q$
according to their Renormalization Group Equations (RGE). The explicit $Q$
dependence on $\widetilde T_i^{\overline{MS}}(Q)$ is cancelled at one--loop
by the implicit $Q$ dependence on the parameters of the tree level tadpoles.
Renormalized tadpoles must be zero at the minimum of the potential
$t_i=0$, thus the generalization of the tadpole equations is
\begin{eqnarray}
&\left[(m_{H_1}^2+\mu^2)v_1-B\mu v_2-\mu\epsilon_3v_3+
\eighth(g^2+g'^2)v_1(v_1^2-v_2^2+v_3^2)\right](Q)+\widetilde 
T_1^{\overline{MS}}(Q) = 0 \,,
\nonumber \\
&\left[(m_{H_2}^2+\mu^2+\epsilon_3^2)v_2-B\mu v_1+
B_2\epsilon_3v_3-\eighth(g^2+g'^2)v_2(v_1^2-v_2^2+v_3^2)\right](Q)
+\widetilde T_2^{\overline{MS}}(Q) = 0 \,,
\nonumber\\
&\left[(m_{L_3}^2+\epsilon_3^2)v_3-\mu\epsilon_3v_1+
B_2\epsilon_3v_2+\eighth(g^2+g'^2)v_3(v_1^2-v_2^2+v_3^2)\right](Q)
+\widetilde T_3^{\overline{MS}}(Q) = 0 \,.
\label{eq:1loopTadpolesEq}
\end{eqnarray}
and these are the minimization condition we impose
\footnote{
To see the effect one--loop tadpoles have on the determination of
MSSM--SUGRA parameters, see ref.~\cite{marcotadpoles}
}. 
We choose to work at the scale $Q=m_Z$. The RGE's for each parameter are 
given in the Appendix A, and the boundary condition at the GUT scale are 
described later.

Now we find the one--loop contributions to the tadpoles. 
Quarks contribute to $\chi^0_1$ and $\chi^0_2$ one--loop tadpoles only. 
On the contrary, squarks contribute to all three tadpoles. Using the notation
for the Feynman rules introduced in the previous section, the quark and squark 
one--loop contribution to the tadpoles can be written as:
\begin{eqnarray}
\left[T_{\chi^0_1}\right]^{tb\tilde t\tilde b}&=&
{{N_c}\over{16\pi^2}}\sum_{i=1}^2
\left[M_{\chi^0_1\tilde t\tilde t}^{ii}A_0(m_{\tilde t_i}^2)
+M_{\chi^0_1\tilde b\tilde b}^{ii}A_0(m_{\tilde b_i}^2)\right]
+{{N_cgm_b^2}\over{8\pi^2m_Wc_{\beta}s_{\theta}}}A_0(m_b^2)
\nonumber\\
\left[T_{\chi^0_2}\right]^{tb\tilde t\tilde b}&=&
{{N_c}\over{16\pi^2}}\sum_{i=1}^2
\left[M_{\chi^0_2\tilde t\tilde t}^{ii}A_0(m_{\tilde t_i}^2)
+M_{\chi^0_2\tilde b\tilde b}^{ii}A_0(m_{\tilde b_i}^2)\right]
+{{N_cgm_t^2}\over{8\pi^2m_Ws_{\beta}s_{\theta}}}A_0(m_t^2)
\nonumber\\
\left[T_{\tilde\nu_{\tau}^R}\right]^{tb\tilde t\tilde b}&=&
{{N_c}\over{16\pi^2}}\sum_{i=1}^2
\left[M_{\tilde\nu_{\tau}^R\tilde t\tilde t}^{ii}A_0(m_{\tilde t_i}^2)
+M_{\tilde\nu_{\tau}^R\tilde b\tilde b}^{ii}A_0(m_{\tilde b_i}^2)\right]
\label{eq:T1T2T3tbstsb}
\end{eqnarray}
where $A_0$ is the first Veltman's function defined by
\begin{equation}
A_0(m^2)=m^2(\Delta-\ln \frac{m^2}{Q^2}+1)
\label{eq:A0}
\end{equation}
The finite tadpoles $\widetilde T_i^{\overline{MS}}(Q)$ are found simply
by setting $\Delta=0$ in the previous expressions.

%%\section{Connection with the GUT Scale}

\section{Unification}

We now discuss the corresponding boundary conditions at unification. 
We assume that at the unification scale the model is characterized 
by one universal soft supersymmetry-breaking mass $m_0$ for all the 
scalars (the gravitino mass), and a
universal gaugino mass $M_{1/2}$. Moreover we assume that there is 
a single trilinear soft breaking scalar mass parameter $A$ and that
the bilinear soft breaking parameter $B$ is related to $A$ through
$B=A-1$. 
In other words we make the standard minimal supergravity assumptions:
\bea
A_t = A_b = A_{\tau} \equiv A \:, 
\\ B=B_2=A-1 \:, \\
m_{H_1}^2 = m_{H_2}^2 = M_{L}^2 = M_{R}^2 = m_0^2 \:, \\
M_{Q}^2 =M_{U}^2 = M_{D}^2 = m_0^2 \:, \\
M_3 = M_2 = M_1 = M_{1/2}
\label{univ}
\eea
at $Q = M_{GUT}$. At energies below $M_{GUT}$ these conditions do not 
hold, due to the renormalization group evolution from the unification 
scale down to the relevant scale.

%\section{Potential}

In order to determine the values of the Yukawa couplings and of the
soft breaking scalar masses at low energies we first run the RGE's from
the unification scale $M_{GUT} \sim 10^{16}$ GeV down to the weak
scale. In doing this we randomly give values at the unification scale
for the parameters of the theory. The range of variation of the MSSM-SUGRA
parameters at the unification scale is as follows
\begin{equation}
\begin{array}{ccccc}
10^{-2} & \leq &{h^2_t}_{GUT} / 4\pi & \leq&1 \cr
10^{-5} & \leq &{h^2_b}_{GUT} / 4\pi & \leq&1 \cr
-3&\leq&A/m_0&\leq&3 \cr
0&\leq&\mu^2_{GUT}/m_0^2&\leq&10 \cr
0&\leq&M_{1/2}/m_0&\leq&5 \cr
\end{array}
\label{unification}
\end{equation}
while the range of variation of $\epsilon_3$ is
\begin{equation}
10^{-2} \leq {\epsilon^2_3}_{GUT}/m_0^2 \leq 10 
\label{epsunification}
\end{equation}
and the value of ${h^2_{\tau}}_{GUT}/ 4 \pi$ is defined in such a way
that we get the $\tau$ mass correctly. After running the RGE we have a 
complete set of parameters, Yukawa couplings and soft-breaking masses 
$m^2_i(RGE)$ to study the minimization.

Similar to what happens in the MSSM-SUGRA (see Appendix B) the number of
independent parameters of this model is actually less than given
above, as one must take into account the W mass constraint as well
as the minimization conditions. In the end there is a single new
parameter characterizing our model, namely $\epsilon_3$.

\section{Results and Phenomenology}

\begin{figure}
\centerline{\protect\hbox{\psfig{file=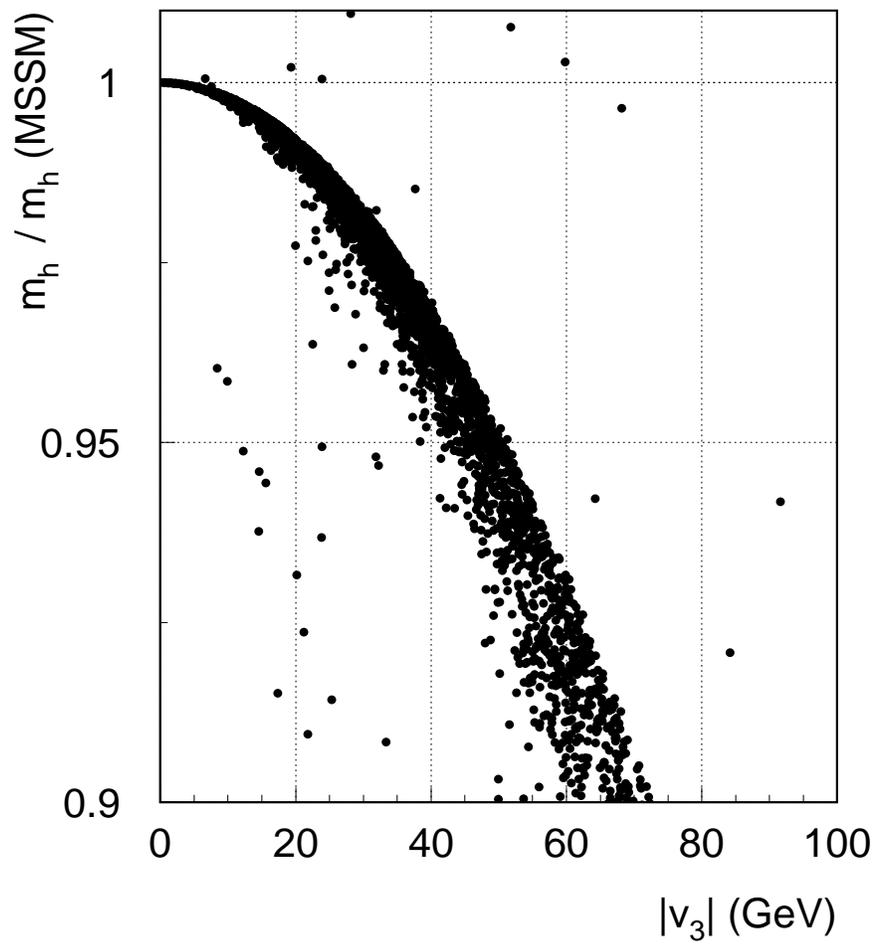,height=13cm,width=0.72\textwidth}}}
\caption{Lightest CP-even Higgs boson mass $m_h$ as a function of
$v_3$ in our model} 
\label{mssmlimit} 
\end{figure}

The main parameters characterizing electroweak breaking are the SU(2)
doublet VEVs $v_1$, $v_2$ and $v_3$. In our model these are obtained
as explained in the Appendix B. Basically we assign random
values for the top and bottom quark Yukawa couplings $h_t$ and $h_b$
at the GUT scale and evolve them down to the weak scale through the
Renormalization Group Equations, given in Appendix A. Using the
measured top and bottom quark masses we determine the corresponding
running masses at the weak-scale.  Combining this with the values of
$h_t$ and $h_b$ at the weak-scale, obtained through the use of the
RGE's, we calculate the standard MSSM VEVS $v_1$ and $v_2$.  The third
VEV $v_3$, which breaks R-parity, is determined through the W mass
formula.  The resulting VEVs may not be consistent with the
minimization conditions. In Appendix B we present a procedure to
ensure a consistent solution. Note that due to the contribution of
$v_3$ to the intermediate gauge boson masses, $v_1^2 + v_2^2$ is
smaller than in the MSSM. The first check of we can do to verify the
consistency of the model is to study the allowed values of the
lightest CP-even Higgs boson mass $m_h$ as a function of the third VEV
$v_3$. This is displayed in Fig.~(\ref{mssmlimit})
The unrotated neutral CP-even Higgs bosons $\chi^0_1$ and $\chi^0_2$
mix with the real part of the tau sneutrino $\tilde\nu^R_{\tau}$. These
are the CP--even scalars $S^0_i$, $i=1,2,3$, introduced in section 3.
The mass matrix can be written as
\begin{equation}
{\bf M}^2_{S^0}={\bf M}^2_{S^0,MSSM}+{\bf M}^2_{S^0,\epsilon_3}+
{\bf M}^2_{S^0,RC}
\label{eq:S0massDec}
\end{equation}
where ${\bf M}^2_{S^0,MSSM}$ is the MSSM mass matrix given by
\begin{equation} 
{\bold M_{S^0,MSSM}^2}=
\left[\matrix{
B\mu{{v_2}\over{v_1}}+\quarter g_Z^2v_1^2
& -B\mu-\quarter g^2_Zv_1v_2 
& 0
\cr -B\mu-\quarter g^2_Zv_1v_2 
& B\mu{{v_1}\over{v_2}}+\quarter g^2_Zv_2^2
& 0
\cr 0
& 0
& m_{L_3}^2+\eighth g^2_Z(v_1^2-v_2^2)
}\right] 
\label{eq:S0massMSSM}
\end{equation}
where we have defined $g_Z^2\equiv g^2+g'^2$.
As expected, this mass matrix has no mixing between the Higgs and stau 
sectors. The extra terms that appear in our $\epsilon_3$--model are
\begin{equation} 
{\bold M_{S^0,\epsilon_3}^2}=
\left[\matrix{
\mu\epsilon_3{{v_3}\over{v_1}}
& 0
& -\mu\epsilon_3+\quarter g^2_Zv_1v_3 
\cr 0
& -B_2\epsilon_3{{v_3}\over{v_2}}
& B_2\epsilon_3-\quarter g^2_Zv_2v_3 
\cr -\mu\epsilon_3+\quarter g^2_Zv_1v_3 
& B_2\epsilon_3-\quarter g^2_Zv_2v_3 
& \epsilon_3^2+{3\over 8} g^2_Zv_3^2
}\right] 
\label{eq:S0massEps3}
\end{equation}
which introduce a Higgs--Stau mixing. Finally, in ${\bf M}^2_{S^0,RC}$
we introduce the largest term in the one--loop radiative corrections,
\ie, the term proportional to $m_t^4$: 
\begin{equation} 
{\bold M_{S^0,RC}^2}=
\left[\matrix{ 0 & 0 & 0
\cr 0 & \Delta_t & 0
\cr 0 & 0 & 0
}\right]\,,\qquad
\Delta_t={{3g^2m_t^4}\over{16\pi^2m_W^2s_{\beta}^2s_{\theta}^2}}\,\ln\,
{{m_{\tilde t_1}^2m_{\tilde t_2}^2}\over{m_t^4}}\,;
\label{eq:S0massRC}
\end{equation}
This formula gives results good in first approximation, nevertheless,
already in the MSSM can give wrong results in certain regions of 
parameter space \cite{marcoalpha}, and should be improved. 

\begin{figure}
\centerline{\protect\hbox{\psfig{file=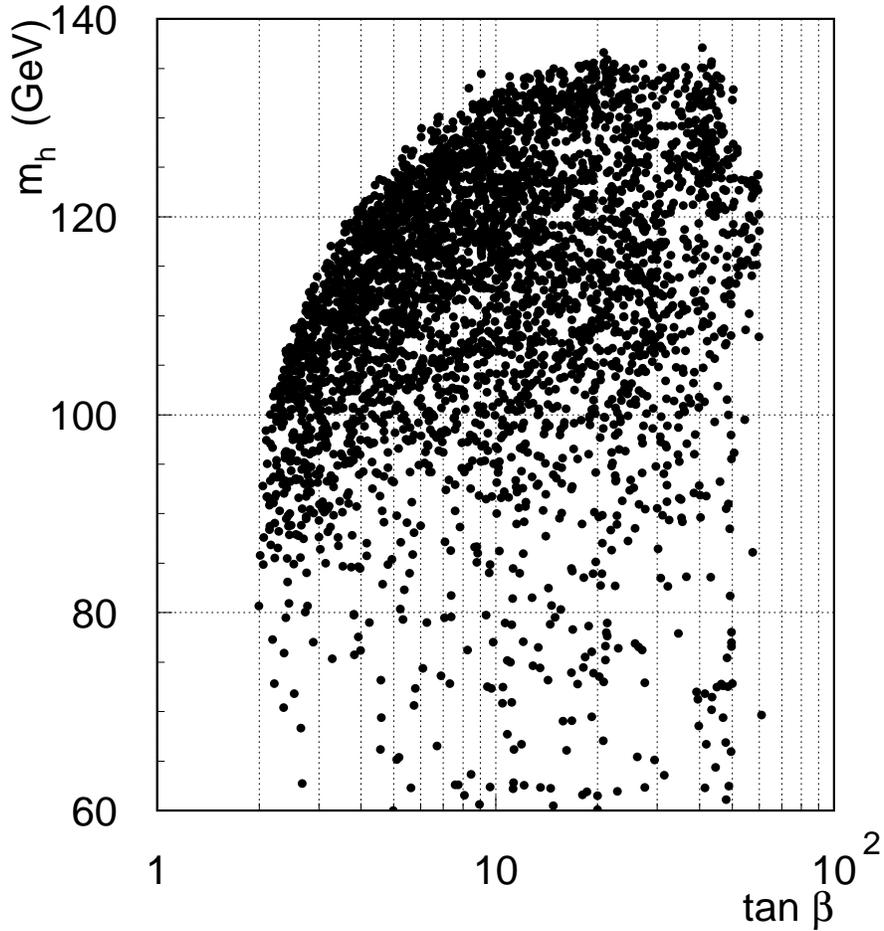,height=13cm,width=0.72\textwidth}}}
\caption{Lightest CP-even Higgs boson mass $m_h$ versus $\tan\beta$}
\label{mh_tanbeta_60} 
\end{figure}

\begin{figure}
\centerline{\protect\hbox{\psfig{file=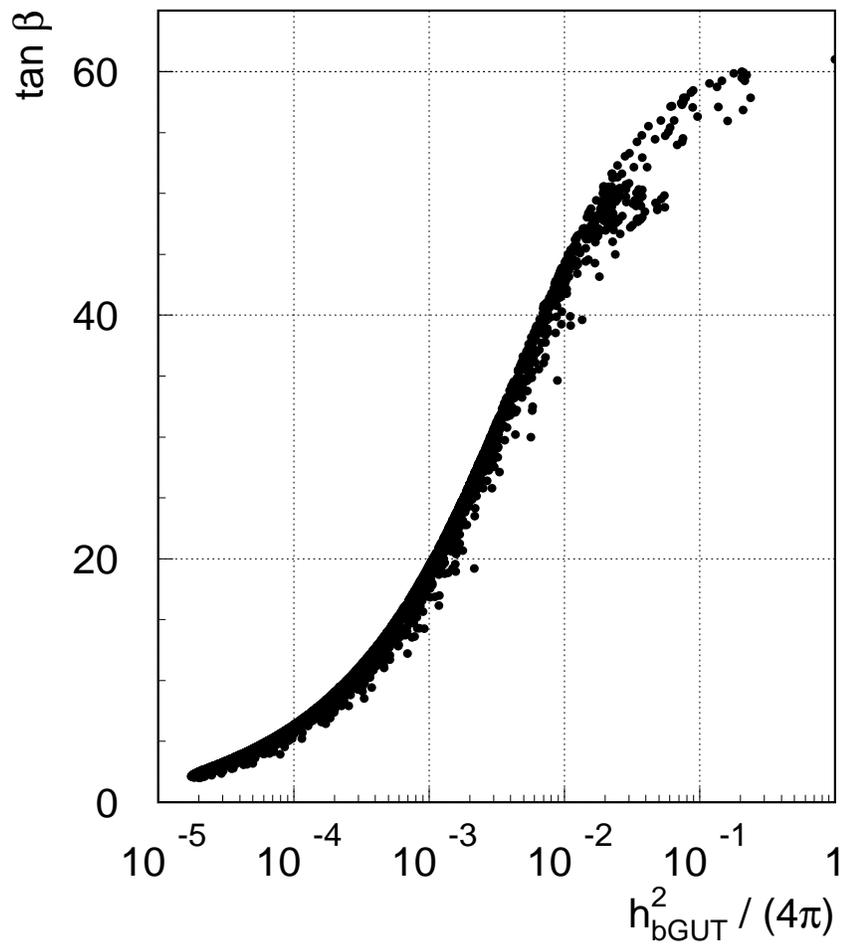,height=13cm,width=0.72\textwidth}}}
\caption{$\tan \beta$ versus bottom quark Yukawa coupling at
unification} 
\label{tanbeta} 
\end{figure}

As one can see in Fig.~(\ref{mssmlimit}), in the limit $v_3 \to 0$ our
model reproduces exactly the expected minimal SUGRA limit for the
lightest CP-even Higgs boson mass. Another view of the Higgs boson
mass spectrum allowed in our model is obtained by plotting $m_h$ as a
function of $\tan\beta$, as illustrated in Fig.~(\ref{mh_tanbeta_60}).
One sees that all values of $\tan\beta$ in the range 2 to 60 or so are
possible in our model. As in the MSSM--SUGRA, $\tan\beta$ smaller than
2 are not possible because the top Yukawa coupling diverges as we
approach the unification scale. This is related to the fact that in
that region we are close to the infrared quasi--fixed point.  
Note that the range of $\tan\beta$ values obtained in our model is
consistent with the unification hypothesis for a large range of the
bottom quark Yukawa coupling at unification, as illustrated in
Fig.~(\ref{tanbeta}).  

\begin{figure}
\centerline{\protect\hbox{\psfig{file=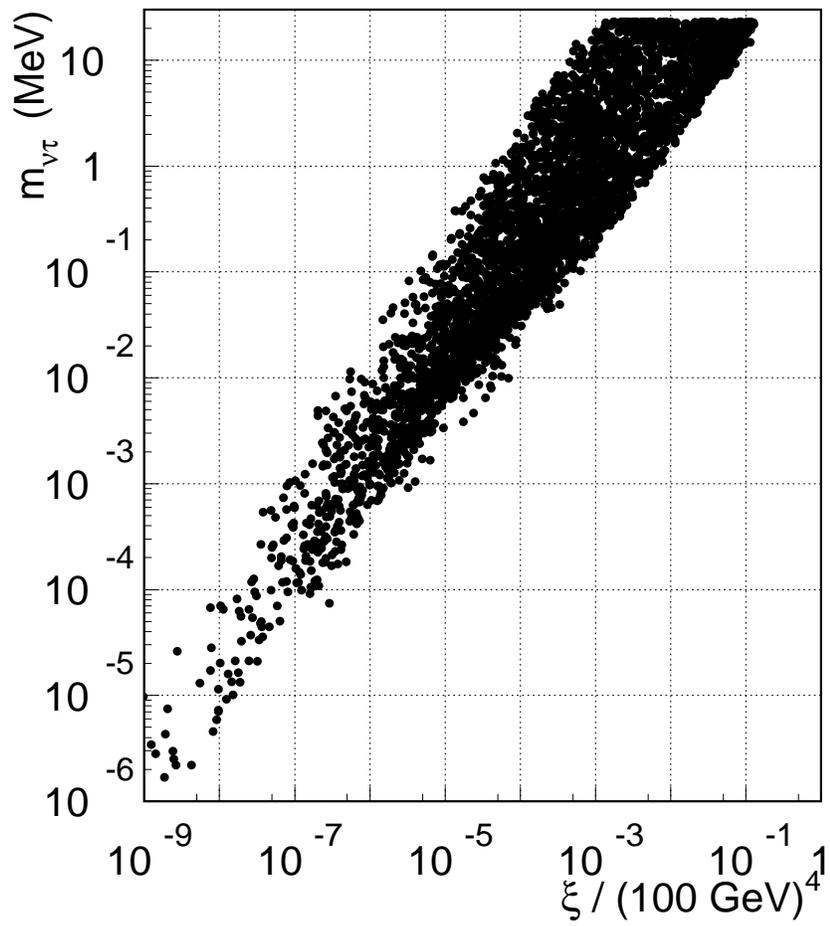,height=13cm,width=0.72\textwidth}}}
\caption{Tau neutrino mass versus $\xi \equiv (\epsilon_3 v_1 + \mu v_3)^2$ }
\label{mnt_xi_ev}
\end{figure}

Another important feature of our broken R-parity model is that the tau
neutrino $\nu_{\tau}$ acquires a mass, due to the fact that $\epsilon_3$ and
$v_3$ are nonzero.  
Consider the basis $(\Psi^0)^T=(-i\lambda_1,-i\lambda_2^3,\widetilde H_1^1,
\widetilde H_2^2, \nu_{\tau})$, where $\lambda_1$ is the $U(1)$ gaugino
introduced in eq.~(\ref{eq:Vsoft}), $\lambda_2^3$ is the neutral $SU(2)$
gaugino, $\widetilde H_i^i$, $i=1,2$ are the neutral Higgsinos, and
$\nu_{\tau}$ is the SM tau neutrino. In this base, the mass terms in the 
Lagrangian for the neutralino--neutrino sector are
\begin{equation}
{\cal L}_m=-\half(\Psi^0)^T{\bf M}_N\Psi^0+h.c.
\label{eq:NeutrMassTerm}
\end{equation}
where the mass matrix is
\footnote{More complete forms of this matrix have
been given in many places. See, e.g. ref. \cite{Romao92}}
\begin{equation}
{\bf M}_N=\left[\matrix{
M_1 & 0  & -\half g'v_1 & \half g'v_2 & -\half g'v_3 \cr
0   & M_2 & \half g v_1 & -\half g v_2 & \half g v_3 \cr
-\half g'v_1 & \half g v_1 & 0 & -\mu & 0 \cr
\half g'v_2 & -\half g v_2 & -\mu & 0 & \epsilon_3 \cr
-\half g'v_3 & \half g v_3 & 0 & \epsilon_3 & 0 
}\right]
\label{eq:NeutMassMat}
\end{equation}
The only new terms appear in the mixing between neutralinos and
tau--neutrino. This mixing is proportional to $\epsilon_3$ and $v_3$.
They lead to a non-zero Majorana $\nu_{\tau}$ mass, which
depends quadratically on the lepton-number-violating parameters
$\epsilon_3$ and $v_3$. Thus R-parity violation in this model
is the origin of neutrino mass. In Fig.~(\ref{mnt_xi_ev}) we
display the allowed values of $m_{\nu_{\tau}}$ (in the tree level
approximation) as a function of an effective parameter $\xi$
defined as $\xi \equiv (\epsilon_3 v_1 + \mu v_3)^2$
\footnote{This combination appears in treating the neutral 
fermion mass matrix in the seesaw approximation.}
Notice that $m_{\nu_{\tau}}$ values can cover a very wide range, from
eV to values in the MeV range, comparable to the present LEP limit
\cite{eps95}.  The latter places a limit on the value of
$\epsilon_3$. Note that the values of $v_3$ and $\epsilon_3$ can 
be rather large [see, for example, Fig.~(\ref{mssmlimit})].

\section{Discussion and Conclusions}

Here we have shown that this simplest truncated version of the
R-parity breaking model of ref. \cite{mv90}, characterized by a
bilinear violation of R-parity in the superpotential, is consistent
with minimal N=1 supergravity models with radiative electroweak
symmetry breaking and universal scalar and gaugino masses at the
unification scale.  We have performed a thorough study of the
minimization of the scalar boson potential of the model, using the
tadpole method.  We have determined the lowest-lying CP-even Higgs
boson mass spectrum.  We have discussed how the minimal N=1
supergravity limit of this theory is obtained and verified that it
works as expected. We have determined also the ranges of $\tan\beta$
and bottom quark Yukawa couplings allowed at unification, as well as
the relation between the tau neutrino mass and the effective bilinear
R-parity violating parameter. Our results should encourage further
theoretical work on this model, as well as more complete versions of
the model, like that of ref. \cite{rprad}. Phenomenological studies of
the related signals should also be desirable, given the fact that the
production and decay patterns of Higgs bosons and supersymmetric
particles in this model are substantially different that expected in
the MSSM or MSSM-SUGRA. For example, Higgs bosons may have sizeable
R-parity violating decays \cite{eps0+}. Similarly, sneutrinos and
staus can be the LSP and can have unsuppressed decays into standard
model states, thus violating R-parity. Finally, chargino and
neutralino production can lead to totally different signals as, for
example, the lightest neutralino can decay \cite{sensi}. These
features could play an important role in designing strategies for
searching for supersymmetric particles at future accelerators.  For
example, R-Parity will give rise to enhanced lepton multiplicities
in Gluino Cascade Decays at LHC \cite{lhc}.

\section*{Acknowledgements}

This work was supported by DGICYT under grants PB95-1077 
and Acci\'on Integrada HP96-055, and by the TMR network 
grant ERBFMRXCT960090 of the European Union. M. A. D. was 
supported by a DGICYT postdoctoral grant.

\newpage
\section*{Appendix A: The Renormalization Group Equations}

Here we will give the renormalization group equations for the
model described by the superpotential in eq.~(\ref{eq:Wsuppot}),
but including only the third generation, and by the soft supersymmetry
breaking terms given in eq.~(\ref{eq:Vsoft}). First we write
the equations for the yukawa couplings of the trilinear terms:
\begin{equation}
16 \pi^2 \frac{d h_U}{d t} =
h_U \left( 6 h_U^2 + h_D^2 - \frac{16}{3} g_3^2 - 3 g_2^2 
- \frac{13}{9} g_1^2 \right) 
\end{equation}
\begin{equation}
16 \pi^2 \frac{d h_D}{d t} =
h_D \left( 6 h_D^2 +  h_U^2+  h_{\tau}^2
- \frac{16}{3} g_3^2 - 3 g_2^2 
- \frac{7}{9} g_1^2 \right) 
\end{equation}
\begin{equation}
16 \pi^2 \frac{d h_{\tau}}{d t} =
h_{\tau} \left( 4 h_{\tau}^2 + 3 h_D^2 
- 3 g_2^2 - 3 g_1^2 
\right) 
\end{equation}
now the corresponding cubic soft supersymmetry breaking
parameters
\begin{equation}
8 \pi^2 \frac{d A_U}{d t} =
6 h_U^2 A_U + h_D^2 A_D 
+ \frac{16}{3} g_3^2 M_3 + 3 g_2^2 M_2
+ \frac{13}{9} g_1^2 M_1 
\end{equation}
\begin{equation}
8 \pi^2 \frac{d A_D}{d t} =
 6 h_D^2 A_D +  h_U^2 A_U +  h_{\tau}^2 A_{\tau}
+ \frac{16}{3} g_3^2 M_3+ 3 g_2^2 M_2 
+ \frac{7}{9} g_1^2 M_1 
\end{equation}
\begin{equation}
8 \pi^2 \frac{d A_{\tau}}{d t} =
 4 h_{\tau}^2 A_{\tau} + 3 h_D^2 A_D 
+ 3 g_2^2 M_2 + 3 g_1^2 M_1 
\end{equation}
For the soft supersymmetry breaking mass parameters we have
\begin{eqnarray}
8 \pi^2 \frac{d M_{Q}^2}{d t} &=&
h_U^2 ( m_{H_2}^2 + M_Q^2 + M_U^2 + A_U^2)
+ h_D^2 ( m_{H_1}^2 + M_Q^2 + M_D^2 + A_D^2) \cr 
&&
- \frac{16}{3} g_3^2 M_3^2 -3 g_2^2 M_2^2 - \frac{1}{9}g_1^2 M_1^2 
+ \frac{1}{6}~g_1^2 {\cal S}
\end{eqnarray}
\begin{equation}
8 \pi^2 \frac{d M_{U}^2}{d t} =
2 h_U^2 ( m_{H_2}^2 + M_Q^2 + M_U^2 + A_U^2)
- \frac{16}{3} g_3^2 M_3^2 - \frac{16}{9}g_1^2 M_1^2
-\frac{2}{3}~g_1^2 {\cal S}
\end{equation}
\begin{equation}
8 \pi^2 \frac{d M_{D}^2}{d t} =
2 h_D^2 ( m_{H_1}^2 + M_Q^2 + M_D^2 + A_D^2)
- \frac{16}{3} g_3^2 M_3^2 - \frac{4}{9}g_1^2 M_1^2
+\frac{1}{3}~g_1^2 {\cal S}
\end{equation}
\begin{equation}
\label{ML}
8 \pi^2 \frac{d M_L^2}{d t} =
h_{\tau}^2 ( m_{H_1}^2 + M_L^2 + M_R^2 + A_{\tau}^2)
-3 g_2^2 M_2^2 - g_1^2 M_1^2
-\frac{1}{2}~g_1^2 {\cal S}
\end{equation}
\begin{equation}
8 \pi^2 \frac{d M_R^2}{d t} =
2 h_{\tau}^2 ( m_{H_1}^2 + M_L^2 + M_R^2 + A_{\tau}^2)
 - 4 g_1^2 M_1^2
+ g_1^2 {\cal S}
\end{equation}
\begin{equation}
8 \pi^2 \frac{d m_{H_2}^2}{d t} =
3 h_U^2 ( m_{H_2}^2 + M_Q^2 + M_U^2 + A_U^2) 
- 3 g_2^2 M_2^2 - g_1^2 M_1^2 
+ \frac{1}{2}~g_1^2 {\cal S}
\end{equation}
\begin{eqnarray}
\label{MHD}
8 \pi^2 \frac{d m_{H_1}^2}{d t} &=&
3 h_D^2 ( m_{H_1}^2 + M_Q^2 + M_D^2 + A_D^2) +
h_{\tau}^2 ( m_{H_1}^2 + M_L^2 + M_R^2 + A_{\tau}^2)\cr 
&&
-3 g_2^2 M_2^2 - g_1^2 M_1^2
- \frac{1}{2}~g_1^2 {\cal S}
\end{eqnarray}
where
\begin{equation}
{\cal S}=  m_{H_2}^2 - m_{H_1}^2 + M_Q^2 -2 M_U^2 + M_D^2 
- M_L^2 + M_R^2 
\end{equation}
For the bilinear terms in the superpotential we get
\begin{equation}
16 \pi^2 \frac{d \mu}{d t} =
\mu \left(  3 h_U^2 + 3 h_D^2 + h_{\tau}^2  
- 3 g_2^2 - g_1^2 
\right) 
\end{equation}
\begin{equation}
16 \pi^2 \frac{d \epsilon_3}{d t} =
\epsilon_3 \left(  3 h_U^2 +  h_{\tau}^2  
- 3 g_2^2 - g_1^2 
\right) 
\end{equation}
and for the corresponding soft breaking terms
\begin{equation}
\label{B}
8 \pi^2 \frac{d B}{d t} =
3 h_U^2 A_U + 3 h_D^2 A_D + h_{\tau}^2 A_{\tau}  
+ 3 g_2^2 M_2 + g_1^2 M_1 
\end{equation}
\begin{equation}
\label{B2}
8 \pi^2 \frac{d B_2}{d t} =
3 h_U^2 A_U + h_{\tau}^2 A_{\tau}  
+ 3 g_2^2 M_2 + g_1^2 M_1 
\end{equation}
The $g_i$ are the $SU(3)\times SU(2)\times U(1)$ gauge couplings and
the $M_i$ are the corresponding the soft 
breaking gaugino masses.

\section*{Appendix B: Minimization Procedure}

To minimize the scalar potential we use the procedure developed in
refs. \cite{rprad,minpot}. We solve the tadpole equations,
eq.~(\ref{eq:1loopTadpolesEq}), for the soft mass-squared parameters
in terms of the VEVS and of the other parameters at the weak
scale. This is particularly simple because those equations are linear
in the soft masses squared. To do this we need to know the values for
the VEVS. These are obtained in the following way:

\begin{enumerate}

\item
We start with random values for $h_t$ and $h_b$ at $M_{GUT}$ in the range
given in eq.~(\ref{unification}). The value of $h_{\tau}$ at $M_{GUT}$
is fixed in order to get the correct $\tau$ mass.

\item
The value of $v_1$ is determined from $m_{b}=h_b v_1/ \sqrt{2}$ for
$m_{b}=3$ GeV (running $b$ mass at $m_Z$). 

\item
The value of $v_2$ is determined from $m_{t}=h_t v_2/ \sqrt{2}$ for
$m_{t}=176 \pm 5$ GeV. If 
\begin{equation}
v_1^2+v_2^2 > v^2=\frac{4}{g^2}\, m^2_W = (246 \hbox{ GeV})^2
\end{equation}
we go back and choose another starting point.

\item
The value of $v_3$ is then obtained from $\displaystyle
v_3=\pm\, \sqrt{\frac{4}{g^2}\, m^2_W -v_1^2 -v_2^2}$.

\end{enumerate}

\noindent
We see that the freedom in $h_{t}$ and $h_{b}$ at $M_{GUT}$ can be
translated into the freedom in the mixing angles $\beta$ and
$\theta$. Comparing, at this point, with the MSSM we have one extra
parameter $\theta$. We will discuss this in more detail below. In 
the MSSM we would have $\theta=\pi/2$.

After doing this, for each point in parameter space, we solve the extremum
equations, eq.~(\ref{eq:1loopTadpolesEq}),  for the soft breaking
masses, which we now call $m^2_i$ ($i=H_1,H_2,L$). 
Then we calculate numerically the eigenvalues for the real and
imaginary part of the neutral scalar mass-squared matrix. If they are
all positive, except for the Goldstone boson, the point is a good one. 
If not, we go back to the next random value. After doing this we end up
with a set of solutions for which:

\begin{enumerate}

\item
The Yukawa couplings are determined by the procedure described above.

\item
The other parameters are given by the RGE evolution once the values at
$M_{GUT}$ are fixed. Notice, however, that these parameters may not 
satisfy the tadpole equations. We will come back to this later.

\item
For a given set of $m^2_i$ ($i=H_1,H_2,L$) each point is also a
solution of the minimization of the potential.

\item
However,  the $m^2_i$ obtained from the minimization
of the potential differ from those obtained from the RGE, which we
call  $m^2_i(RGE)$. 

\end{enumerate}

Our next goal is to find which solutions, for the $m^2_i$ that minimize the
effective low-energy potential, have the property that they 
coincide with the $m^2_i(RGE)$ obtained, for a given unified theory, 
from the RGE, namely
\begin{equation}
m^2_i=m^2_i(RGE) \quad ; \quad i=H_1,H_2,L
\end{equation}
Following ref. \cite{rprad} we define a function
\begin{equation}
\eta= \max \left( \frac{m^2_i}{m^2_i(RGE)},\frac{m^2_i(RGE)}{m^2_i}
\right) \quad ; \quad \forall i
\end{equation}
Defined in this way it is easy to see that we always have $\eta \geq 1$,
the equality being what we are looking for. 

We are then all set for a minimization procedure. We want, by varying
the parameters at the GUT scale, to get $\eta$ as close to 1 as
possible.  With these conditions we used the {\tt MINUIT} package in
order to find the minimum of $\eta$. We considered a point in
parameter space to be a good solution if $\eta < 1.001$.

Before we end this Appendix, let us discuss the counting of free
parameters in this model and in the minimal N=1 supergravity unified
version of the MSSM.  As we explained above after requiring the
correct masses for the $W$, $t$, $b$ and $\tau$ we get one free
parameter in the MSSM, $\tan \beta$, and two in our model, $\tan
\beta$ and $\cos \theta$ or, equivalently, $v_3$. As for the other
parameters we have at the GUT scale one extra parameter,
$\epsilon_3$. But we also have an extra equation for the tadpoles. So in
the end our model has just one more free parameter. This can be
summarized in the following tables:

\begin{center}
\begin{tabular}{|c|c|c|}\hline
Parameters & Conditions & Free Parameters \cr \hline
$h_t$, $h_b$, $h_{\tau}$, $v_1$, $v_2$
&$m_W$, $m_t$, $m_b$, $m_{\tau}$ & $\tan
\beta$ \cr \hline
$A$, $m_0$, $M_{1/2}$, $\mu$&$t_i=0$, $i=1,2$& 2 Extra free
parameters\cr \hline
Total = 9&Total = 6 &Total = 3\cr\hline
\end{tabular}

{Table 1: Counting of free parameters in minimal N=1 supergravity}
\end{center}

\begin{center}
\begin{tabular}{|c|c|c|}\hline
Parameters & Conditions & Free Parameters \cr \hline
$h_t$, $h_b$, $h_{\tau}$, $v_1$, $v_2$, $v_3$
&$m_W$, $m_t$, $m_b$, $m_{\tau}$ & $\tan\beta$, $\cos \theta$ \cr \hline
$A$, $m_0$, $M_{1/2}$, $\mu$, $\epsilon_3$
&$t_i=0$, $i=1,2,3$& 2 Extra free parameters \cr \hline
Total = 11&Total = 7 &Total = 4\cr\hline
\end{tabular}

{Table 2: Counting of free parameters in our model}
\end{center}

Finally, we note that in either case, the sign of the mixing parameter
$\mu$ is physical and has to be taken into account.

\end{document}